\documentclass{article}

\usepackage{arxiv}

\usepackage[utf8]{inputenc} % allow utf-8 input
\usepackage[T1]{fontenc}    % use 8-bit T1 fonts
\usepackage{hyperref}       % hyperlinks
\usepackage{url}            % simple URL typesetting
\usepackage{booktabs}
\usepackage{multirow}
\usepackage{subcaption}% professional-quality tables
\usepackage{amsfonts}       % blackboard math symbols
\usepackage{nicefrac}       % compact symbols for 1/2, etc.
\usepackage{microtype}      % microtypography
\usepackage{lipsum}		% Can be removed after putting your text content
\usepackage{graphicx}
\usepackage{natbib}

\title{Influential Prototypical Networks for Few Shot Learning: A Dermatological Case Study}

\author{ \href{https://orcid.org/0000-0003-3706-3227}{\includegraphics[scale=0.06]{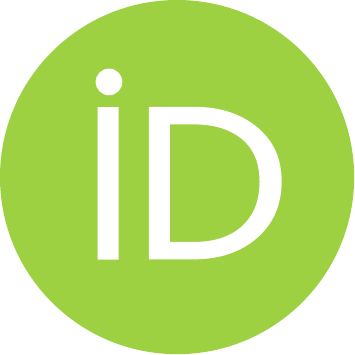}\hspace{1mm}Ranjana Roy Chowdhury}\\
	Department of Computer Science \& Engineering\\ 
	Indian Institute of Technology Ropar, India\\
	\texttt{ranjana.20csz0002@iitrpr.ac.in} \\
	\And
	\href{https://orcid.org/0000-0002-1383-3744}{\includegraphics[scale=0.06]{orcid.pdf}\hspace{1mm}Deepti R. Bathula} \\
	Department of Computer Science \& Engineering\\ 
	Indian Institute of Technology Ropar, India\\
	\texttt{bathula@iitrpr.ac.in} \\
}

\begin{document}
\maketitle

\begin{abstract}
Prototypical network (PN) is a simple yet effective few shot learning strategy. It is a metric-based meta-learning technique where classification is performed by computing Euclidean distances to prototypical representations of each class. Conventional PN attributes equal importance to all samples and generates prototypes by simply averaging the support sample embeddings belonging to each class. In this work, we propose a novel version of PN that attributes weights to support samples corresponding to their influence on the support sample distribution. Influence weights of samples are calculated based on maximum mean discrepancy (MMD) between the mean embeddings of sample distributions including and excluding the sample. Comprehensive evaluation of our proposed influential PN (IPNet) is performed by comparing its performance with other baseline PNs on three different benchmark dermatological datasets. IPNet outperforms all baseline models with compelling results across all three datasets and various $N$-way, $K$-shot classification tasks. Findings from cross-domain adaptation experiments further establish the robustness and generalizability of IPNet.
\end{abstract}

% keywords can be removed
\keywords{Prototypical Networks \and Few Shot Learning \and Influence Factor \and Maximum Mean Discrepancy (MMD)}

\section{Introduction}
Over the last decade, rapid evolution of deep learning algorithms has led to applications in many diverse areas including medicine. Their resounding efficacy is demonstrated through state-of-the-performance across wide variety of computer vision tasks. However, these algorithms require large datasets with diverse examples to learn from. Unfortunately, prevalence of such large annotated datasets is quite uncommon in fields like medical imaging. To avoid issues of overfitting with small datasets, several alternatives have been proposed including transfer learning and domain adaptation.

Recently, Few Shot Learning (FSL), a type of meta-learning has been proposed that aims to learn from just a few examples like humans. While several variants of FSL have been designed that leverage prior knowledge about similarity, data and learning \citep{Yaqing}, metric-based Prototypical Networks (PN) have gained significant attention due their simplicity and efficiency. These networks learn a metric space where classification is performed by computing Euclidean distances to prototypical representations of each class. In standard PN \citep{Jake}, prototype of each class is generated by simply averaging the support samples belonging to that class in the latent feature space. This naïve approach considers each support sample to be equally important for prototype formation.

Since their inception, many versions of PNs have been proposed to improve their performance. In \citep{Kushagra}, Group Equivariant convolutions (G-convolutions) are incorporated into PN to compensate for lack of canonical structure in dermatological images. A re-weighing mechanism is explored in \citep{Junjie}, which reduces the influence of noisy samples when learning the prototype. This is achieved by an adaptive strategy of inverse distance weighting of samples to the feature embedding learned on the remaining samples. These efforts lack the potential to explicitly weight the support samples on the basis of their distinguishing characteristics i.e., samples that are similar (prototypes) and those that are dissimilar (criticisms) to the prototypes \citep{Been} . \citep{Rushil} proposed a method to identify influential samples by weighting them based on Maximum Mean Discrepancy (MMD). Inspired by this work, we propose a simple variation of classical PN that attributes weights to support samples based on their MMD score that represents their influence on the support sample distribution. The class-wise prototypes are then formed by taking the weighted mean of the corresponding support samples in the latent feature space. Towards this end, the main contributions of our work are as follows:

\begin{itemize}
    
    \item We propose a novel version of prototypical network that assesses the influence of each sample based on the MMD score between the mean embeddings of sample distributions including and excluding the sample.
    
    \item Support samples are assigned influence weights, inverse of MMD weights, to form the prototypical class representations in the embedded space.

    \item We demonstrate the efficacy of the proposed Influential Prototypical Network (IPNet) by applying it to three benchmark dermatological datasets and compare with other versions of PNs.

    \item Additionally, we establish the robustness and generalizability of IPNet through cross-domain experiments.
    
\end{itemize}

\section{METHODOLOGY}
\label{sec:method}
\subsection{Preliminaries}
\label{ssec:prelim}
Here we have considered the task based episodic training process of FSL for our problem formulation. For each training episode, random samples are chosen from $N$ classes to create a support set $S=(\textbf{s}_{i},y_{i})^{K}_{i=1}$ and query set $Q=(\textbf{q}_{i},y_{i})^{K}_{i=1}$. Here, $\textbf{s}_{i}$ and $\textbf{q}_{i}$ represent the sampled images and $y_{i}$ their corresponding labels belonging to $N$ categories. As the support set contains $K$ training samples, this represents the $N$-way, $K$-shot classification task.

The support set forms the crux of prototype formation in PN.  The classic PN is a metric based FSL approach that learns a metric space by mapping each input $\textbf{x}_{i} \in \mathbb{R}^{D}$ by an embedding function $f_{\theta}(\textbf{x}_{i}): \mathbb{R}^{D} \longrightarrow \mathbb{R}^{M} $ that is parameterized by $\theta$. In each episode, it computes a prototypical representation of each class by taking the mean of all support sample embeddings of that class as:

\begin{equation}
\label{eqn:proto}
\textbf{p}_{c}=\frac{1}{|S_{c}|}\sum_{\textbf{x}_{s}\in{S_{c}}}f_{\theta}(\textbf{x}_{i})
\end{equation}

\noindent where $S_{c}$ is the set of samples from class $c$. Subsequently, the class label of a new sample from the query set $Q$ is predicted by calculating the Euclidean distance of the sample $\textbf{x}_{i}$ to each class prototypical vector and applying softmax on the distances as:

\begin{equation}
\label{eq2}
p_{\theta}(y_{i} = c\ |\ \textbf{x}_{i}, \textbf{p}_{c}) = \frac{ \exp(-d(f_{\theta}(\textbf{x}_{i}),\textbf{p}_{c}))}{\sum_{c'} \exp (-d(f_{\theta}(\textbf{x}_{i}),\textbf{p}_{c'}))}
\end{equation}

\noindent where $d(.)$ is the Euclidean distance function between query sample and prototypical vector. The $\theta$ parameter is updated likewise in order to improve the likelihood computed on $Q$ and is given as:

\begin{equation}
\label{eq3}
\sum_{(x_{j},y_{j})\in{Q}}\log{ p(y_{i} = c|\textbf{x}_{i})}
\end{equation}

\noindent where $y_{i}$ is the ground truth of $\textbf{x}_{i}$. Although simple and quite effective, this  naïve  approach  considers  each  support sample to be equally important for prototype formation.

The recently introduced RRPNet \citep{Junjie} replaces the simple arithmetic mean used for computing the prototypes with a weighted average of the support sample embeddings to reduce the influence of noisy samples. Their adaptive weighting scheme assigns weight to each sample based on its distance to the prototype learned from the remaining support sample embeddings. 

\subsection{Proposed Approach}
\label{ssec:proposed}

Inspired by the efforts on influential sample selection \citep{Rushil} using MMD \citep{Ilya}, we propose a novel and principled approach to creating prototypes. The idea is to assign weights to the samples according to their influence on the sample distribution of that class. And the influence of a particular sample can be measured by how much the distribution changes in the absence of that sample. We use MMD for this purpose.

MMD is a kernel based approach that measures the distributional discrepancy between two datasets as distance between the mean embeddings of their features. Given two datasets with distributions $A$ and $B$ respectively, the MMD between them is given as:

\begin{equation}
\label{eqn:mmd1}
\textsf{MMD}_{\phi}(A,B) = ||\mu_{\phi}(A)-\mu_{\phi}(B)||
\end{equation}

\noindent where $\phi$ represents the mapping function to the latent space. Consequently, the conformity of a particular data sample ($\textbf{s}$) to its corresponding dataset distribution ($V$) can be measured using MMD as:

\begin{equation}
\label{eqn:mmd2}
\textsf{MMD}(\textbf{s}) = \textsf{MMD}_{\phi}(V,V') = ||\mu_{\phi}(V)-\mu_{\phi}(V')||
\end{equation}

\noindent where $V$ represents the whole dataset and $V'$ is the same dataset but excluding the sample $\textbf{s}$. As $\textsf{MMD}_{\phi}(A,B) = 0\ \  \textrm{iff}\ \  A = B$, samples with lower MMD scores indicate high compliance with the distribution and data points with high MMD score signify deviation from the sample distribution. As the samples with high conformity to the distribution should be given more importance in creating the prototypical vectors, we define the influential (IF) weight of sample as $\textsf{IF}(\textbf{s}) = 1\  - \  \textsf{MMD}(\textbf{s})$ after normalizing the MMD scores of all samples in the support set.

As a result, the prototypical representation of each class in our proposed IPNet is formed using:

\begin{equation}
\label{eqn:ifproto}
\textbf{p}_{c} = \frac{\sum^{|S_{c}|}_{i=1} \textsf{IF}(f_{\theta}(\textbf{x}_{i}))f_{\theta}(\textbf{x}_{i})}{\sum^{|S_{c}|}_{i=1} \textsf{IF}(f_{\theta}(\textbf{x}_{i}))}
\end{equation}
\large
\begin{table*}
%\scalebox{0.6}{

\resizebox{\columnwidth}{!}{ 
\renewcommand{\arraystretch}{2}
\begin{tabular}{|l||l|l||l|l|l|l||l|l|l|l|}
\hline
& \multicolumn{2}{c||}{\large{ISIC-2018}} &   \multicolumn{4}{c||}{\large{Derm7pt}}     & \multicolumn{4}{c|}{\large{SD-198}}       
\\ \hline
& \multicolumn{2}{c||}{\large{2-way}} &        \multicolumn{2}{c|}{\large{2-way}}    & \multicolumn{2}{c||}{\large{5-way}}   & \multicolumn{2}{c|}{\large{2-way}}    & \multicolumn{2}{c|}{\large{5-way}}    
\\ \hline
& \begin{tabular}[c]{@{}l@{}}\large{3-shot}\end{tabular} & \begin{tabular}[c]{@{}l@{}}\large{5-shot}\end{tabular} & \begin{tabular}[c]{@{}l@{}}\large{3-shot}\end{tabular} & \begin{tabular}[c]{@{}l@{}}\large{5-shot}\end{tabular} & \begin{tabular}[c]{@{}l@{}}\large{3-shot}\end{tabular} & \begin{tabular}[c]{@{}l@{}}\large{5-shot}\end{tabular} & \begin{tabular}[c]{@{}l@{}}\large{3-shot}\end{tabular} & \begin{tabular}[c]{@{}l@{}}\large{5-shot}\end{tabular} & \begin{tabular}[c]{@{}l@{}}\large{3-shot}\end{tabular} & \begin{tabular}[c]{@{}l@{}}\large{5-shot}\end{tabular} \\ \hline

PNet \citep{Jake}  & 65.52 (0.71) &  74.21 (0.75)                                & 61.23 (0.64) &  67.21 (0.69)                                & 60.23 (0.62) &  63.21 (0.66)                                & 67.21 (0.70) &  74.22 (0.78)                                & 66.21 (0.69) &  71.69 (0.76)                 \\ \hline
MDDNet \citep{Kushagra} & 73.50 (0.76) & 79.70 (0.83)                                 & 66.80 (0.65) & 69.50 (0.73)                                 & -             & -                                           & 72.10 (0.75) & 80.20 (0.86)                                 & -             & -                          \\ \hline
RRPNet \citep{Junjie} & 76.32 (0.79) & 80.25 (0.84)                                 & 72.10 (0.74) & 77.49 (0.80)
                 & 70.36 (0.67) & 75.56 (0.75)                & 74.89 (0.76) & 80.23 (0.86)                & 74.36 (0.75) & 77.90 (0.82)    
\\ \hline
IPNet (ours) & \textbf{79.00 (0.83)} & \textbf{84.20 (0.87)}               & \textbf{75.66 (0.79)} & \textbf{80.21 (0.86)}               & \textbf{75.35 (0.77)} & \textbf{78.39 (0.82)}               & \textbf{78.41 (0.83)} & \textbf{84.20 (0.87)}               & \textbf{77.23 (0.81)} & \textbf{81.44 (0.84)}  \\ \hline
\end{tabular}
\renewcommand{\arraystretch}{1}
}
\caption{Performance comparison for intra-domain classification: Average accuracy and AUC (in brackets) for various $N$-way, $K$-shot classification tasks across three benchmark dermatological datasets.}
\label{table:intra}
\end{table*}

\section{Experimental Results}
\label{sec:experiments}

\subsection{Datasets}
\label{ssec:data}

To evaluate the performance of our proposed IPNet on the few-shot classification task in the medical domain, we chose three benchmark dermatological datasets including ISIC-2018 \citep{Noel}, Dermp7pt \citep{Jeremy} and SD-198 \citep{Xiaoxiao}. From these three datasets, we utilized classes with maximum number of images as training classes (4, 13 \& 20) and classes with very less count representing rare skin disease categories as test classes (3, 6 \& 70) respectively. We also applied standard data augmentation (including rotation, scaling, etc.) to generate relatively balanced training classes. 

\subsection{Implementation Details}
\label{ssec:impl}

To ensure fair comparison, we use an architecture similar to the one proposed in \citep{Kushagra}, the standard Conv-6 backbone with batch normalization. It is a 6 layer CNN where each block consisting of a $3 \times 3$ convolutional layer with $64$ channels and a $2 \times 2$ max-pooling layer followed by Stochastic Gradient Descent (SGD) for optimization and ReLU as the activation function. We use a batch size of $5$ and set the learning rate and momentum to 0.01and 0.9 respectively. Furthermore, we implemented the standard PN from \citep{Jake} (PNet) and robust, re-weighting PN from \citep{Junjie} (RRPNet) using the same framework described above for objective comparison with our approach across various few-shot classification tasks. We also compare with the reported results of \citep{Kushagra} (MDDNet) on the same datasets.

\subsection{Experiments}
\label{ssec:exp}

We analyze both performance and robustness of different versions of prototypical networks in the context of (a) intra-domain, few-shot classification of unseen classes and (b) cross-domain adaptation with training and testing data belong to different domains as described below.

\subsubsection{Intra-domain Classification}
\label{sssec:intra}

For this experiment, the training and testing classes are chosen from the same dataset to represent common diseases with lot of samples and rare diseases with very few samples respectively. Here, we perform $2$-way and $5$-way classification tasks on each of the three datasets where $N$-way refers to number of classes randomly chosen from the training classes with train-shot of ten images per class for training the models. Similarly, at test time, $N$ classes are chosen randomly from the unseen test classes with $3$ and $5$ support samples from each class to form $3$-shot and $5$-shot classification tasks.  For ISIC-2018 dataset, the $5$-way classification task is omitted as it has only $3$ test classes. The average accuracy (along with standard deviation) and AUC values across $2000$ testing episodes are used for performance analysis. These results are summarized in Table \ref{table:intra}.

\subsubsection{Cross-domain Adaptation}
\label{sssec:cross}

To evaluate the robustness as well as generalizability of the proposed IPNet, we perform cross-domain validation task where different versions of the prototypical models are trained on one dataset and tested on the remaining datasets. Here, we permute among the three datasets to ensure that each datasets gets assigned as training domain at least once. Similar to the intra-domain analysis, we experiment with both $3$-shots and $5$-shots but limit the evaluation to only $2$-way classification tasks as the ISIC-2018 contains only three test classes. Table \ref{table:cross} depicts the results of cross-domain analysis using average accuracy and AUC as the metrics 

\subsection{Results}
\label{ssec:results}

\begin{figure}
     \centering
     \begin{subfigure}[b]{0.3\textwidth}
         \centering
         \includegraphics[width=\textwidth]{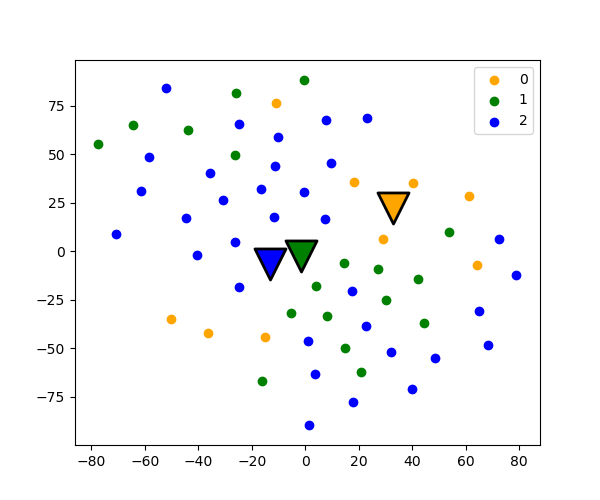}
         \caption{PNet}
         
     \end{subfigure}
     \hfill
     \begin{subfigure}[b]{0.3\textwidth}
         \centering
         \includegraphics[width=\textwidth]{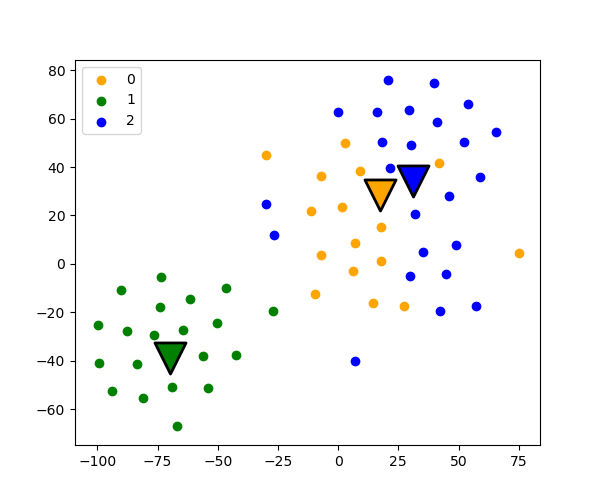}
         \caption{RRPNet}
         
     \end{subfigure}
     \hfill
     \begin{subfigure}[b]{0.3\textwidth}
         \centering
         \includegraphics[width=\textwidth]{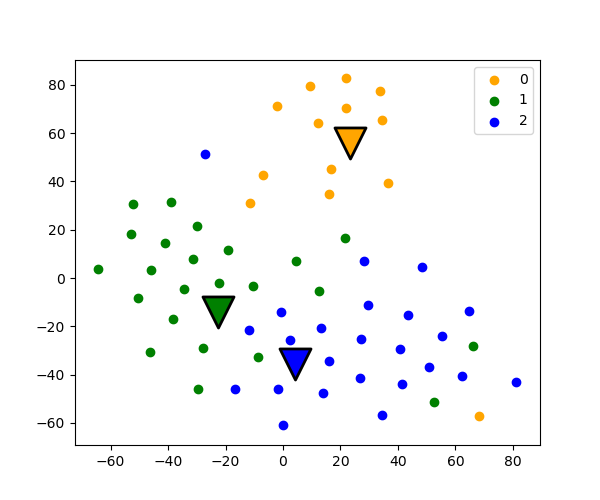}
         \caption{IPNet}
         
     \end{subfigure}
        \caption{The t-SNE visualization of prototype representation on Derm7pt dataset for 3-Way Classification Task.}
        \label{fig:fig1}
\end{figure}

\begin{figure}
     \centering
     \begin{subfigure}[b]{0.3\textwidth}
         \centering
         \includegraphics[width=\textwidth]{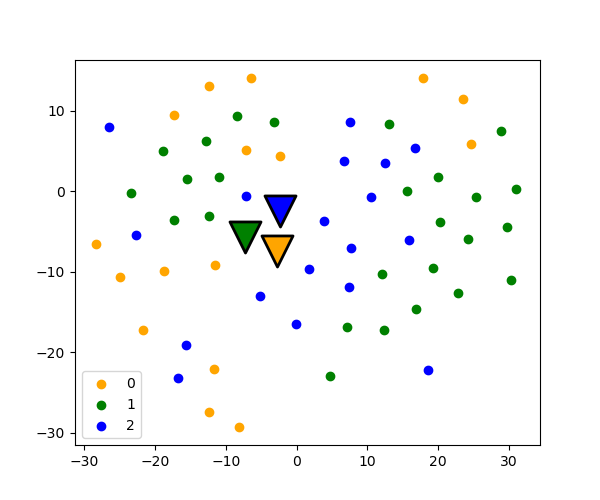}
         \caption{PNet}
         
     \end{subfigure}
     \hfill
     \begin{subfigure}[b]{0.3\textwidth}
         \centering
         \includegraphics[width=\textwidth]{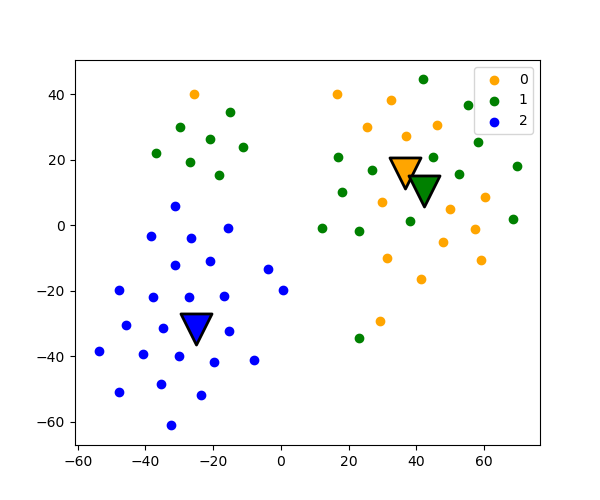}
         \caption{RRPNet}
         
     \end{subfigure}
     \hfill
     \begin{subfigure}[b]{0.3\textwidth}
         \centering
         \includegraphics[width=\textwidth]{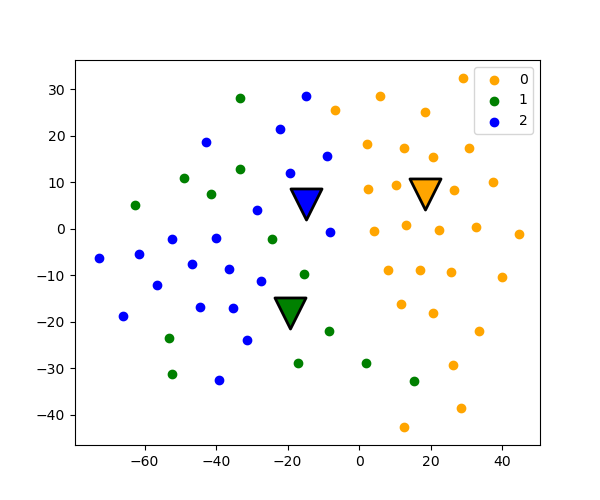}
         \caption{IPNet}
         
     \end{subfigure}
        \caption{The t-SNE visualization of prototype representation on SD-198 dataset for 3-Way Classification Task.}
        \label{fig:fig2}
\end{figure}

As highlighted in Table \ref{table:intra}, our proposed IPNet outperforms all other baseline prototypical networks – PNet, MDDNet and RRPNet across all three dermatological datasets and various $N$-way, $K$-shot classification tasks.
On average, IPNet provides an improvement of $3.5\%$ and $6.4\%$ in average accuracy and $5.1\%$ and $7.8\%$ in AUC when compared to RRPNet and MDDNet respectively. Similar trends are also observed in Table \ref{table:cross} for cross-domain analysis. For $2$-way classification on test domains, with both $3$-shots and $5$-shots, IPNet provides the best classification accuracy and AUC as compared to PNet and RRPNet. Finally, qualitative comparison of prototype representations of the models using t-SNE plots is shown in Figures \ref{fig:fig1} and \ref{fig:fig2} for Derm7pt and SD-198 datasets respectively.  IPNet provides distinct clustering of support samples with clear segregation of class-specific prototypes. %compared to other models.

%In comparison to other baselines, IPNet provides improvement in average accuracy and AUC values ranging from approximately $2.5\%$ to $6\%$ for intra-domain, FSL classification tasks.

\begin{table}
%\scalebox{0.55}{
\resizebox{\columnwidth}{!}{
\renewcommand{\arraystretch}{2}
\begin{tabular}{|l|l|l|l|l|l|l|l|l|l|l|l|l|l|}
\hline
                                                                                  & \textbf{TRAIN}  & \multicolumn{4}{c|}{\textbf{Derm7pt}}                                          & \multicolumn{4}{c|}{\textbf{ISIC-2018}}                                      & \multicolumn{4}{c|}{\textbf{SD-198}}                                            \\ \hline
                                                                                  & \textbf{TEST}   & \multicolumn{2}{c|}{\textbf{SD-198}} & \multicolumn{2}{c|}{\textbf{ISIC-2018}} & \multicolumn{2}{c|}{\textbf{SD-198}} & \multicolumn{2}{c|}{\textbf{Derm7pt}} & \multicolumn{2}{c|}{\textbf{Derm7pt}} & \multicolumn{2}{c|}{\textbf{ISIC-2018}} \\ \hline
\multirow{3}{*}{\textbf{\begin{tabular}[c]{@{}l@{}}2-Way \\ 3-Shot\end{tabular}}} & \textbf{PNet\citep{Jake}}   & \multicolumn{2}{l|}{$58.44\pm0.18\ (0.60)$}                & \multicolumn{2}{l|}{$60.32\pm0.19\ (0.61)$}                   & \multicolumn{2}{l|}{$60.23\pm0.20\ (0.65)$}                & \multicolumn{2}{l|}{$61.23\pm0.18\ (0.63)$}                 & \multicolumn{2}{l|}{$62.77\pm0.17\ (0.68)$}                 & \multicolumn{2}{l|}{$61.98\pm0.18\ (0.67)$}                   \\ \cline{2-14} 
                                                                                  & \textbf{RRPNet\citep{Junjie}} & \multicolumn{2}{l|}{$69.80\pm0.17\ (0.71)$}                & \multicolumn{2}{l|}{$70.12\pm0.18\ (0.71)$}                   & \multicolumn{2}{l|}{$68.10\pm0.19\ (0.70)$}                & \multicolumn{2}{l|}{$67.50\pm0.17\ (0.70)$}                 & \multicolumn{2}{l|}{$70.00\pm0.10\ (0.72)$}                 & \multicolumn{2}{l|}{$68.32\pm0.18\ (0.70)$}                   \\ \cline{2-14} 
                                                                                  & \textbf{IPNet}  & \multicolumn{2}{l|}{$\textbf{71.25}\pm\textbf{0.17\  (0.74)}$}                & \multicolumn{2}{l|}{$\textbf{72.60}\pm\textbf{0.18\  (0.75)}$}                   & \multicolumn{2}{l|}{$\textbf{71.32}\pm\textbf{0.18\  (0.73)}$}                & \multicolumn{2}{l|}{$\textbf{70.21}\pm\textbf{0.17\ (0.74)}$}                 & \multicolumn{2}{l|}{$\textbf{73.11}\pm\textbf{0.17\  (0.75)}$}                 & \multicolumn{2}{l|}{$\textbf{71.10}\pm \textbf{0.17}\  \textbf{(0.74)}$}                   \\ \hline\hline
\multirow{3}{*}{\textbf{\begin{tabular}[c]{@{}l@{}}2-Way\\ 5-Shot\end{tabular}}}  & \textbf{PNet\citep{Jake}}   & \multicolumn{2}{l|}{$63.66\pm0.17\ (0.64)$}                & \multicolumn{2}{l|}{$63.11\pm0.18\ (0.65)$}                   & \multicolumn{2}{l|}{$67.22\pm0.19\ (0.70)$}                & \multicolumn{2}{l|}{$66.90\pm0.17\ (0.71)$}                 & \multicolumn{2}{l|}{$68.9\pm0.17\ (0.71)$}                 & \multicolumn{2}{l|}{$67.2\pm0.17\ (0.72)$}                   \\ \cline{2-14} 
                                                                                  & \textbf{RRPNet\citep{Junjie}} & \multicolumn{2}{l|}{$72.80\pm0.17 \  (0.76)$}                & \multicolumn{2}{l|}{$71.23\pm0.18 \  (0.75)$}                   & \multicolumn{2}{l|}{$72.45\pm0.18 \  (0.74)$}                & \multicolumn{2}{l|}{$71.63\pm0.17 \  (0.74)$}                 & \multicolumn{2}{l|}{$74.38\pm0.17 \  (0.76)$}                 & \multicolumn{2}{l|}{$71.23\pm0.17 \ (0.75)$}                   \\ \cline{2-14} 
                                                                                  & \textbf{IPNet}  & \multicolumn{2}{l|}{$\textbf{76.54}\pm\textbf{0.16}\ \textbf{(0.80)}$}                & \multicolumn{2}{l|}{$\textbf{75.22}\pm\textbf{0.17}\ \textbf{(0.80)}$}                   & \multicolumn{2}{l|}{$\textbf{76.58}\pm\textbf{0.17}\ \textbf{(0.79)}$}                & \multicolumn{2}{l|}{$\textbf{74.12}\pm\textbf{0.16}\ \textbf{(0.77)}$}                 & \multicolumn{2}{l|}{$\textbf{77.11}\pm\textbf{0.16}\ \textbf{(0.79)}$}                 & \multicolumn{2}{l|}{$\textbf{75.21}\pm\textbf{0.16}\ \textbf{(0.78)}$}                   \\ \hline
\end{tabular}
\renewcommand{\arraystretch}{1}
}
\break
\caption{Performance comparison for cross-domain adaption: Average accuracy (with standard deviation) and AUC (in brackets) for models trained on one dataset and tested on the other two.}
\label{table:cross}
\end{table}


\subsection{Conclusion}
\label{ssec:conc}
We proposed a novel version of PN with adaptive weighting scheme that assigns weights to the samples according to their influence on the sample distribution of that class. Further, the influence factor of a sample is measured using MMD based on the shift in the distribution in the absence of that sample. Extensive experiments conducted on three benchmark dermatological datasets and various $N$-way, $K$-shot classification tasks, demonstrate that IPNet achieves compelling results as compared to other versions of PN. Our findings also establish the superiority of IPNet for cross-domain adaptation.

% \usepackage{multirow}

\bibliographystyle{unsrtnat}
\bibliography{arXiv_IPNet1}  %%% Uncomment this line and comment out the ``thebibliography'' section below to use the external .bib file (using bibtex) .





% \end{thebibliography}

\end{document}